\documentclass[preprint,showpacs,aps,prc,superscriptaddress,floatfix]{revtex4-1}
\usepackage{graphicx}

\begin{document}

\title{Strong Multi-step Interference Effects in $^{\mathbf{12}}$C(d,p) to the 9/2$_{\mathbf{1}}^{\mathbf{+}}$ State in $^{\mathbf{13}}$C} 

\author{N. Keeley}
\email{nicholas.keeley@ncbj.gov.pl}
\affiliation{National Centre for Nuclear Research, ul.\ Andrzeja So\l tana 7, 05-400 Otwock, Poland}
\author{K. W. Kemper}
\affiliation{Department of Physics, Florida State University, Tallahassee, Florida 32306, USA}
\author{K. Rusek}
\affiliation{Heavy Ion Laboratory, University of Warsaw, ul.\ Pasteura 5a, 02-093, Warsaw, Poland} 

\begin{abstract}
The population of the 9.50 MeV $9/2_1^+$ resonance in $^{13}$C by single neutron transfer reactions is 
expected to be dominated by the two-step route through the $^{12}$C $2_1^+$ (4.44 MeV) state, with another
possible contribution via the strongly excited $3_1^-$ (9.64 MeV) resonance in $^{12}$C. However, we find that a
good description of the angular distribution for population of this state via the $^{12}$C(d,p)$^{13}$C
reaction is only possible when both direct $0_1^+ \otimes \mathrm{g}_{9/2}$ and two-step (via the 4.44 MeV
$^{12}$C $2_1^+$ state) $2_1^+ \otimes \mathrm{d}_{5/2}$ paths are included in a coupled reaction channel
calculation. While the calculated angular distribution is almost insensitive to the presence of the two-step
path via the 9.64 MeV $^{12}$C $3_1^-$ resonance, despite a much greater contribution to the wave function from the 
$3_1^- \otimes \mathrm{f}_{7/2}$ configuration, its inclusion is required to fit the details of the experimental
angular distribution. The very large interference between the various components of the calculations, even when
these are small, arises through the ``kinematic'' effect associated with the different transfer routes.
\end{abstract}

\pacs{24.50.+g, 25.45.Hi, 24.10.Eq}

\maketitle
\section{Introduction}
The strong population of the very narrow ($\leq 5$ keV) 9.50 MeV $9/2_1^+$ resonance by the $^{12}$C(d,p)$^{13}$C 
reaction provides a means to test two-step reaction models. The observed population is incompatible with a simple
$0_1^+ \otimes \mathrm{g}_{9/2}$ single-particle structure for this level, and structure calculations, see e.g.\ Refs.\
\cite{Ohn85,Mil89}, predict a dominant contribution from the $2_1^+ \otimes \mathrm{d}_{5/2}$ configuration built on the
4.44 MeV $2^+$ excited state of the $^{12}$C core. However, the wave function of this state is expected to contain
a small (of the order of 1 \%) contribution from the $0_1^+ \otimes \mathrm{g}_{9/2}$ configuration as well as larger
components built on the $4_1^+$ excited state of the $^{12}$C core. To date, possible contributions from configurations
built on the 9.64 MeV $3^-$ state of the $^{12}$C core have not been considered. The initial motivation of this work was
therefore to search for the influence of hypothetical components of this type on the measured $^{12}$C(d,p)$^{13}$C angular distribution
for stripping to the $9/2_1^+$ resonance at an incident deuteron energy of 30 MeV \cite{Ohn85}. 

The coupled channels Born approximation (CCBA) analysis of Ohnuma {\em et al.\/} \cite{Ohn85} employed calculated $^{13}$C 
wave functions including components built on the $0_1^+$, $2_1^+$ and $4_1^+$ states of the $^{12}$C core. The two-step (d,p) transitions
proceeding via the $4_1^+$ state were, however, neglected, since they were found to have only a small effect on the results. 
In this work we adopted a different approach, instead varying the spectroscopic amplitudes for the various components
in order to obtain the best description of the angular distribution. We included two-step transfer paths via the $2_1^+$ and
$3_1^-$ states of $^{12}$C as well as the direct path. Following Ohnuma {\em et al.\/} we omitted transfer paths
proceeding via the $^{12}$C $4_1^+$ state. We limited the configurations built on the excited states of the
$^{12}$C core to the $2_1^+ \otimes \mathrm{d}_{5/2}$ and $3_1^- \otimes \mathrm{f}_{7/2}$, since we considered these to be the
most important components, to keep the number of directly variable parameters within manageable proportions. 

The structure of this paper is as follows. Firstly we describe in detail the calculations, then compare the results 
with the high quality data of Ref.\ \cite{Ohn85}, and finally present our conclusions.

\section{Calculations}
All calculations were performed with the code {\sc Fresco} \cite{Tho88}. Inelastic excitation of the 4.44 MeV $2_1^+$ and 9.64 
MeV $3_1^-$ states of $^{12}$C was included as well as the $^{12}$C(d,p)$^{13}$C stripping reaction leading to the 9.50 MeV
$9/2_1^+$ state. The entrance channel optical potential was based on the global parameters of Daehnick {\em et al.\/} \cite{Dae80}
with the real and imaginary well depths adjusted to obtain the best fit to the elastic scattering data of Perrin {\em et al.\/}
\cite{Per77} when the inelastic couplings were included. The $B(E2; 0^+ \rightarrow 2^+)$ and $B(E3; 0^+ \rightarrow 3^-$) were
taken from Refs.\ \cite{Ram01} and \cite{Kib02}, respectively.  Nuclear deformation lengths were fixed by fitting the 60.6 MeV inelastic
scattering data of  Aspelund {\em et al.\/} \cite{Asp75}, yielding values of $\delta_2 = -1.40$ fm and $\delta_3 = 0.65$ fm. 
The exit channel p + $^{13}$C optical potential was calculated using the CH89 global parameters \cite{Var91}. More modern optical
potential parameterizations, e.g.\ that of An and Cai \cite{An06} for the deuteron or that of Koning and Delaroche \cite{Kon03}
for the proton in the exit channel, could equally well have been used as a basis for our calculations. However, we do not expect
that their use would significantly change our results, since Ohnuma {\em et al.\/} \cite{Ohn85} tested several sets of deuteron and
proton potentials in their analysis with no significant difference in the final results. 

The $\left<\mathrm{d}|\mathrm{n} + \mathrm{p}\right>$ overlap was calculated using the Reid soft core potential \cite{Rei68} and
included the small D-state component. Again, a more modern nucleon-nucleon potential such as the Argonne $v_{18}$ \cite{Wir95}
could have been used but the Reid potential is available in {\sc Fresco} as a standard input and we do not anticipate any
significant difference in our results from the use of a more modern parameterization. For the 
$\left<^{13}\mathrm{C}|^{12}\mathrm{C} + \mathrm{n}\right>$ overlaps the neutron was
bound in a Woods-Saxon well with geometry parameters $r_0 = 1.25$ fm, $a_0 = 0.65$ fm. Since the transferred neutron is unbound with respect to
the $^{12}$C core in both its ground and 4.44 MeV $2_1^+$ excited states the $0_1^+ \otimes \mathrm{g}_{9/2}$ and
$2_1^+ \otimes \mathrm{d}_{5/2}$ components of the wave function were calculated in continuum bins, of widths 2.0 and 0.1 MeV,
respectively, the well depths being adjusted to give resonances at the corresponding energies. The transferred neutron is bound
with respect to the $^{12}$C core in its 9.64 MeV $3_1^-$ state so for the $3_1^- \otimes \mathrm{f}_{7/2}$  component of the wave
function the well depth was adjusted to give the appropriate binding energy. 

Initial calculations used the CCBA formalism, i.e.\ the inelastic scattering steps in the entrance channel were treated using
full coupled channel (CC) theory but the transfer steps were described using the distorted wave Born approximation (DWBA). We
varied the spectroscopic amplitudes for the $0_1^+ \otimes \mathrm{g}_{9/2}$, $2_1^+ \otimes \mathrm{d}_{5/2}$ and
$3_1^- \otimes \mathrm{f}_{7/2}$  components of the wave function for the $^{13}$C 9.50 MeV $9/2_1^+$ state to obtain the best description
of the 30 MeV $^{12}$C(d,p)$^{13}$C stripping data of Ohnuma {\em et al.\/} \cite{Ohn85}. However, we found that in order
to describe the data completely it was essential to use full coupled reaction channels (CRC) theory for the transfer
steps. Therefore, the results presented here are for CRC calculations. All calculations were full finite-range CRC and included
both the complex remnant term and the non-orthogonality correction. The post form of the formalism was used in all cases. 

\section{Results}
\begin{figure}
\includegraphics[width=12cm,clip=]{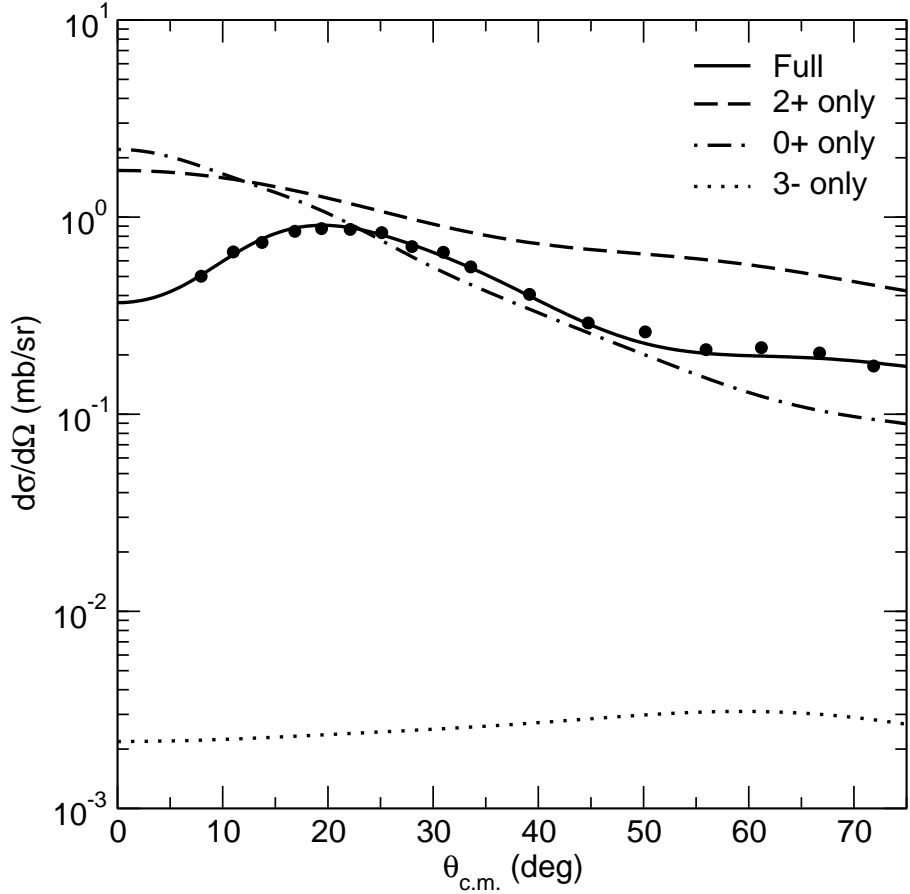}
\caption{\label{fig1}CRC calculations for the $^{12}$C(d,p)$^{13}$C
stripping reaction to the 9.50 MeV $9/2^+$ resonance of $^{13}$C at $E_\mathrm{d} = 30$ MeV compared
with the data of Ref.\ \cite{Ohn85}. The solid curve denotes the full calculation, the dashed curve a calculation including just the
two-step transfer via the $^{12}$C 4.44 MeV $2^+$ state, the dot-dashed curve a calculation including just the direct, one-step
transfer via the $^{12}$C $0^+$ ground state and the dotted curve a calculation including just the two-step transfer via the $^{12}$C
9.64 MeV $3^-$ state.}
\end{figure}

In Fig.\ \ref{fig1} we compare the results of our best fit CRC calculations with the data of Ohnuma {\em et al.\/} \cite{Ohn85}.
The full calculation describes the data well over their whole angular range. A breakdown of the full calculation into its three
components, i.e. direct one-step transfer via the $^{12}$C $0_1^+$ ground state, two-step transfer via the 4.44 MeV $^{12}$C
$2_1^+$ state and two-step transfer via the 9.64 MeV $3_1^-$ state, shows that the shape of the measured angular distribution
can only be satisfactorily reproduced as a result of interference effects between the direct one-step transfer and the two-step
transfer via the $2_1^+$ state. Two-step transfer via the $3_1^-$ state is seen to be almost two orders of magnitude smaller
than the data, making a small but nonetheless necessary contribution to the overall description of the full angular range of the
data.  

The best fit spectroscopic amplitudes for the three components of the 9.50 MeV $9/2_1^+$ wave function are given in Table
\ref{tab1}. While we do not expect these to be {\em quantitatively} accurate---our calculations do not by any means
include all the processes that may influence the absolute values, in particular deuteron breakup---they should be {\em qualitatively}
reasonable, which is sufficient for our purposes here. Including additional effects such as the deuteron breakup would have 
made searching on all three spectroscopic amplitudes prohibitively time consuming.
\begin{table}
\caption{\label{tab1}Best fit spectroscopic amplitudes and corresponding mixing ratios obtained from the analysis of the 
30 MeV $^{12}$C(d,p)$^{13}$C stripping data \cite{Ohn85}.}
\begin{ruledtabular}
\begin{tabular}{c c c}
Component & Spectroscopic amplitude & Mixing ratio (in \%) \\
\hline
$0_1^+ \otimes \mathrm{g}_{9/2}$ & $-0.17$ & 1.57 \\
$2_1^+ \otimes \mathrm{d}_{5/2}$ & $1.33$ & 93.0 \\
$3_1^- \otimes \mathrm{f}_{7/2}$ & $0.32$ & 5.40 \\
\end{tabular}
\end{ruledtabular}
\end{table}
However, we can definitely state that, while small, the spectroscopic amplitude for the $0_1^+ \otimes 
\mathrm{g}_{9/2}$ component must be negative
in order to produce the required interference effect ({\sc Fresco} adopts the convention that all bound state 
wave functions lead off positive from $r=0$).

The corresponding mixing ratios (defined as the percentage contributions of the individual components of the wave function to the whole
in terms of the spectroscopic {\em factors}) show that in spite of the comparable importance of the cross sections the direct one-step
transfer path probes only about $1.6$ \% of the wave function of the $^{13}$C $9/2_1^+$ state compared to $93$ \% for the
two-step path via the $^{12}$C $2_1^+$ excited state. In fact, our mixing ratios compare rather well with those of 
Ohnuma {\em et al.\/} \cite{Ohn85}. We therefore confirm the results of the CCBA calculations of
Ohnuma {\em et al.\/} \cite{Ohn85} who also found that inclusion of the direct one-step transfer path had an
important influence on the $^{12}$C(d,p)$^{13}$C($9/2^+_1$) stripping angular distribution, in spite of the small 
contribution of the $0_1^+ \otimes \mathrm{g}_{9/2}$ component to the $^{13}$C $9/2^+_1$ wave function. 

\section{Discussion of Results}
\begin{figure}
\includegraphics[width=12cm,clip=]{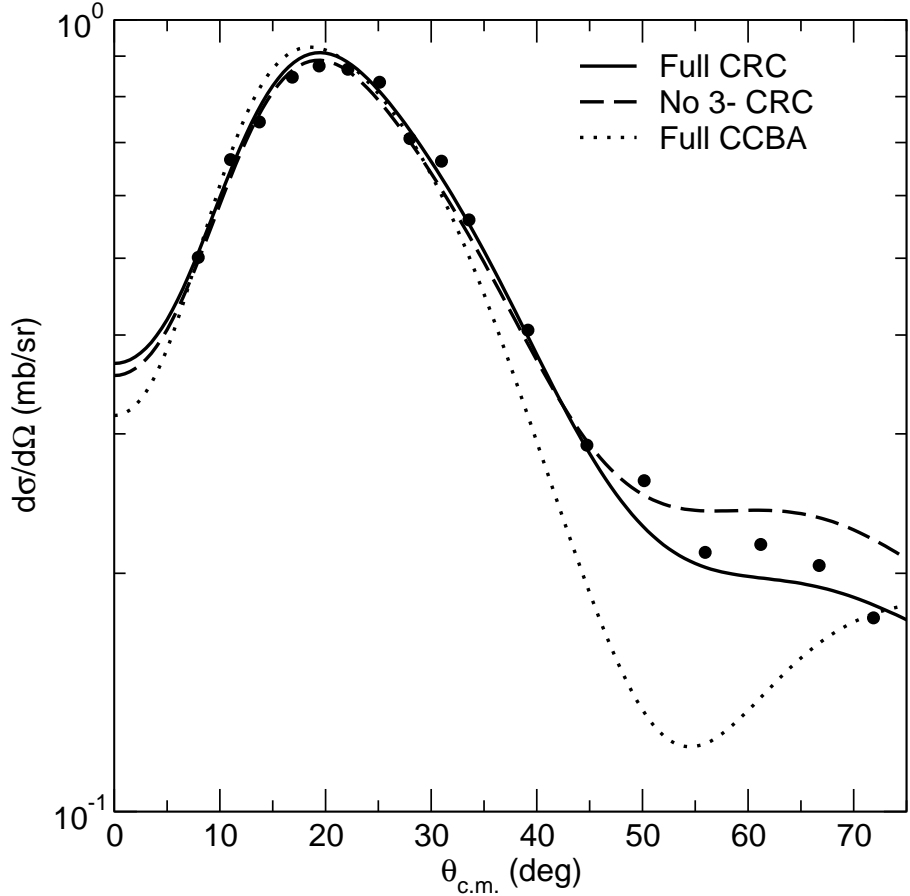}
\caption{\label{fig2}Calculations for the $^{12}$C(d,p)$^{13}$C
stripping reaction to the 9.50 MeV $9/2^+$ resonance of $^{13}$C at $E_\mathrm{d} = 30$ MeV compared
with the data of Ref.\ \cite{Ohn85}. The solid curve denotes the full CRC calculation and the dashed curve a CRC calculation 
omitting the two-step transfer via the 9.64 MeV $^{12}$C $3_1^-$ state. The dotted curve denotes the best-fit full calculation
using CCBA.} 
\end{figure}

We find that the use of CRC as opposed to CCBA improves considerably the agreement between calculated and measured 
angular distributions, eliminating the minimum in the CCBA angular distribution at about $50^\circ$, not
present in the data, see Fig.\ \ref{fig2}. A similar result for CCBA was obtained by Ohnuma {\em et al.\/} \cite{Ohn85}
(cf.\ Fig.\ 9 of Ref.\ \cite{Ohn85}). The main result of our analysis is in many respects similar to 
that of Ohnuma {\em et al.\/} \cite{Ohn85}, since we find that inclusion of the direct one-step transfer path is 
essential to describe the stripping angular distribution, in spite of the small contribution of the 
$0_1^+ \otimes \mathrm{g}_{9/2}$ component to the wave function of the $^{13}$C 
$9/2_1^+$ state. In addition, we find that the small contribution from the two-step path via the $^{12}$C $3_1^-$ state
is necessary for a good description of the entire angular range of the data, see Fig.\ \ref{fig2}; this result has some important 
general implications for other studies of this kind which have not been explored.

The $^{12}$C(d,p)$^{13}$C stripping leading to the 9.50 MeV $9/2_1^+$ resonance in $^{13}$C is a striking reminder that
when attempting to extract nuclear structure information from a fit to direct reaction data the measured cross section
is a product of both nuclear structure factors and what we may term the ``kinematics'' of the reaction, e.g.\ matching
of Q-value and momentum transfer etc. While the addition of the components built on the $4_1^+$ excited state of the
$^{12}$C core had an important influence on the structure calculations, Ohnuma {\em et al.\/} \cite{Ohn85} found their 
contribution to the calculated angular distribution to be negligible. This may be explained as due to the ``kinematic''
effect of having to find the 14 MeV of excitation energy for reaction paths proceeding via the $4_1^+$ core
state and its weak excitation in inelastic scattering, suggesting that it proceeds via a two-step mechanism \cite{Ray78}. 
Even the more modest 9.64 MeV of the $3_1^-$ state is sufficient to damp out almost completely the contribution
of reaction paths proceeding via this state, see Fig. \ref{fig2}. 

Very precise data---with an uncertainty of better than $\pm 10$ \%---are required in order to be sensitive directly to the
contribution from the two-step transfer path via the $^{12}$C $3_1^-$ excited state. We therefore wish to underline that the
``kinematics'' may set a limitation on the sensitivity of direct reaction data to the composition of the wave function of the 
final state in systems where the reaction mechanism has significant contributions from two-step paths. Conversely, due to
the folding of kinematic effects together with the nuclear structure, the inclusion of reaction paths probing apparently 
small components of the wave function may be essential in order to describe the reaction data. The data for deuteron
stripping to the 9.50 MeV $9/2_1^+$ resonance in $^{13}$C currently under discussion provide a  good example of this.

To sum up, when attempting to extract ``empirical'' mixing ratios for the various possible configurations of 
states with suspected significant core excitation components or when trying to validate calculated wave functions 
for such states by comparisons with direct reaction data, it should be borne in mind that kinematics may dominate over structure in the
calculated angular distributions. This is particularly so when the core states are relatively
high-lying; exactly how high-lying will probably depend on the incident energy of the particle inducing the reaction. 
One may well find that under these circumstances the reaction data are actually more sensitive to the smallest components
of the wave function if these happen to be those involving the core nucleus in its ground state or lowest-lying excited
state or states.
\acknowledgments
KWK acknowledges the support of the Robert O. Lawton Fund of Florida State University in this work.

\end{document}